\documentclass[]{aastex7}

\usepackage{verbatim}
\usepackage{amsmath}
\usepackage{graphicx}
\usepackage{url}
\usepackage{footmisc}

\newcommand\Tmag{$T$}
\newcommand\eg{{\it e.g.}}

\begin{document}

\title{QLP Data Release Notes 004: TESS-Gaia Light Curve Photometry Implementation}


\author[0009-0005-2122-0680]{Glen Petitpas}
\affiliation{MIT Kavli Institute for Astrophysics and Space Research, Massachusetts Institute of Technology, Cambridge, MA 02139, USA}
\email{petitpas@mit.edu}

\author[0009-0005-3397-2046]{Jack Haviland}
\affiliation{MIT Kavli Institute for Astrophysics and Space Research, Massachusetts Institute of Technology, Cambridge, MA 02139, USA}
\email{haviland@mit.edu}

\author[0000-0002-7127-7643]{Te Han}
\affiliation{Department of Physics \& Astronomy, The University of California, Irvine, Irvine, CA 92697, USA}
\email{teh2@uci.edu}

\author[0000-0003-0241-2757]{Willie Fong}
\affiliation{MIT Kavli Institute for Astrophysics and Space Research, Massachusetts Institute of Technology, Cambridge, MA 02139, USA}
\email{willfong@mit.edu}

\author[0000-0002-2135-9018]{Katharine Hesse}
\affiliation{MIT Kavli Institute for Astrophysics and Space Research, Massachusetts Institute of Technology, Cambridge, MA 02139, USA}
\email{khesse@mit.edu}

\author[0000-0002-1836-3120]{Avi Shporer}
\affiliation{MIT Kavli Institute for Astrophysics and Space Research, Massachusetts Institute of Technology, Cambridge, MA 02139, USA}
\email{shporer@space.mit.edu}

\author[0000-0002-4371-3460]{Jeroen Audenaert}
\affiliation{MIT Kavli Institute for Astrophysics and Space Research, Massachusetts Institute of Technology, Cambridge, MA 02139, USA}
\email{jeroena@mit.edu}

\author[0000-0002-5788-9280]{Daniel Muthukrishna}
\affiliation{MIT Kavli Institute for Astrophysics and Space Research, Massachusetts Institute of Technology, Cambridge, MA 02139, USA}
\email{danmuth@mit.edu}

\author[0000-0001-6763-6562]{Roland Vanderspek}
\affiliation{MIT Kavli Institute for Astrophysics and Space Research, Massachusetts Institute of Technology, Cambridge, MA 02139, USA}
\email{roland@space.mit.edu}

\author[0000-0003-2058-6662]{George~R.~Ricker}
\affiliation{MIT Kavli Institute for Astrophysics and Space Research, Massachusetts Institute of Technology, Cambridge, MA 02139, USA}
\email{grr@mit.edu}

\begin{abstract}
The Quick-Look Pipeline \cite[QLP;][and references therein]{Huang2020,Kunimoto2021} generates light curves for up to 2 million stars every 27.4 days observed by TESS as part of its planet search. 
As machine learning methods enable deeper searches and scientific priorities shift toward fainter stars, there is a motivation for QLP to perform better at fainter magnitudes. We have adopted the photometry methods employed by the TESS-Gaia Light Curve package \citep{Han2023}, which has been shown to have better noise characteristics than the original QLP photometry from 10.5 $<$ \Tmag\ $<$ 13.5. We still perform aperture photometry and deliver 3 apertures, and 3 levels of detrending for all stars brighter than \Tmag \ = 13.5, so the changes should be seamless for external users. This method is implemented as of Sector 94 in QLP light curves and is providing users with higher precision light curves and allows detection of fainter signals in our planet searches.

\end{abstract}


\section{Introduction}

The Quick-Look Pipeline (QLP), written and operated by the MIT TESS Science Office (TSO) processes data from TESS Full-Frame Images (FFIs), from the raw FFIs to vetting reports for new TESS transiting planet candidates. QLP performs multi-aperture photometry to extract light curves for all targets in the FFIs brighter than TESS magnitude (\Tmag) = 13.5 mag. When processing for a sector is complete, the TSO delivers the new multi-aperture light curves to MAST, and performs a multi-sector transiting planet search for transiting exoplanets combining the new data with all available archival light curve data for each observed target. Our team compiles reports for the high probability candidates, which are then delivered to vetters at the TSO as part of the TOI alert process. For a full description of the QLP light curve extraction procedure, see \cite{Huang2020}. 

\section{Updates}

Initially, due to the amount of human inspection required for signals in the TOI process, QLP photometry and planet search were optimized for targets with \Tmag\ $\leq10.5$. As machine learning efforts allow deeper searches, we are making efforts to improve our photometric precision down to our lightcurve generation limit of \Tmag\ = 13.5. To accommodate this, we have replaced our existing photometry method \citep[FIPHOT;][]{pal2012} with the new TESS-Gaia Light Curve method \citep[TGLC;][]{Han2023}.
The TGLC method uses Gaia DR3 star positions and magnitudes as priors, then forward models the FFIs with the effective point spread function (ePSF) to remove contamination from nearby stars. Details of the process are given in \cite{Han2023}. 

\begin{figure*}
\centering
\includegraphics[width=6.0in]{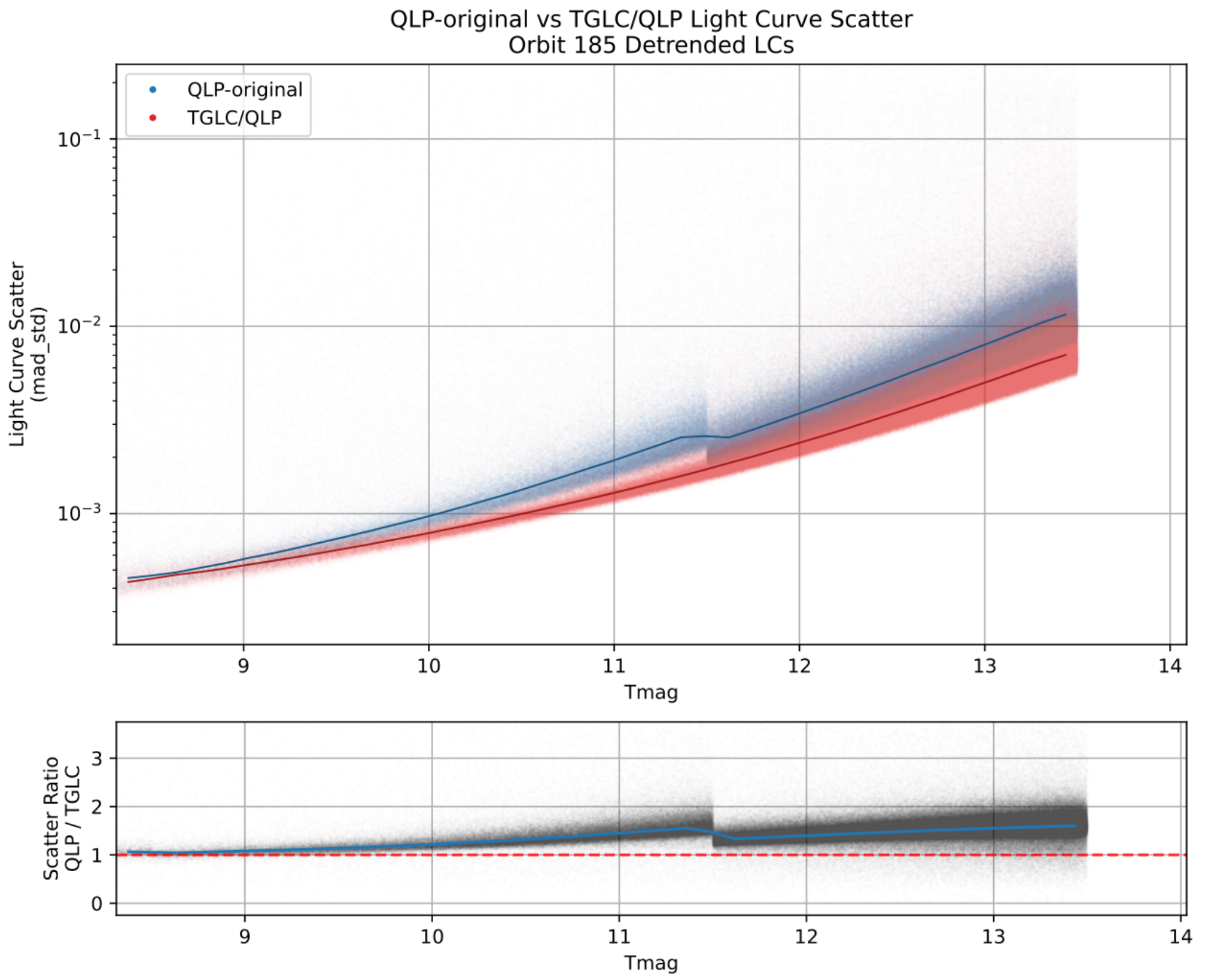}
\caption{Photometric precision [$1.48/\sqrt{2}$ $\times$ (median difference between adjacent flux points), \citet{Han2023}] over the QLP magnitude range for 950,000 light curves from Orbit-185, binned to 30 minute time resolution. The FIPHOT data used in QLP initially is shown in blue, and the updated TGLC values are shown in orange. The lower plot shows the ratio of lightcurve scatter between the two methods. The discontinuity in QLP-original at \Tmag=11.5 is the result of the transition to the smallest fixed QLP aperture.
}
\label{fig:qlpvstglc}
\end{figure*}

Although the TGLC aperture fluxes only use one aperture of $3\times3$ pixel, in order to maintain backward compatibility and enable existing QLP planet search methods to continue (\eg, depth-aperture tests\footnote{Testing for an increase in the transit candidate depth with increasing aperture size, which is the signature of a false positive scenario where the target is contaminated by a nearby stellar eclipsing binary.}), we added two extra apertures to regular TGLC aperture fluxes. The previous version of QLP included 5 aperture sizes and the ``best'' aperture was chosen based on the star brightness. The transitions occurred at \Tmag\ of 6.5, 8.5 and 11.5. The 11.5 transition is visible in Figure \ref{fig:qlpvstglc}. The ``small'' and ``large'' apertures were just one aperture size smaller and larger than the ``best'' \citep{Huang2020}.

  
Now with TGLC the contamination is better controlled, so the main photometry is always extracted with the TGLC's $3\times3$ pixel aperture across the full Tmag range. We added $1\times1$ and $5 \times 5$ pixel apertures for the use of depth-aperture tests and in keeping with the delivery of light curves based on 3 apertures to the public.



\section{Performance}

Despite additional steps such as partitioning the FFIs and deriving ePSFs for each section, the Python routines are run on multiple GPUs and the entire photometry process is complete in about half the time required by our old method. 

Our planet search uses a GPU based Box-Least-Squared (BLS) routine \citep[][]{Kunimoto2023} which is then filtered using a variety of signal-to-noise based cutoffs. Those results are then passed through our machine learning classifier \citep[\texttt{Astronet-Triage};][]{Tey2023} to find eclipse-like signals. Given that our BLS cut-offs were derived from the older data, and \texttt{Astronet-Triage} was trained on lightcurves using the original photometric method, we ensured the improved noise characteristics did not significantly impact our planet search. We ran a series of tests using multi-sector lightcurves generated using a combination of older non-TGLC data plus the newer TGLC generated data. 


Using the original non-TGLC data up to Sector-89 there were $\approx360,000$ TCEs, identified in the light curves of around 955,000 stars, that passed multisector BLS filtering, and of those, 4890 passed \texttt{Astronet-Triage} filtering. Repeating this search, but using new TGLC data for Sector-89 (plus existing non-TGLC data for previous sectors), there were 454 signals that originally passed, but did not pass with the addition of TGLC data. These signals were inspected, and almost all of them were dropped during the BLS noise threshold filtering, in that the original SNR was within a fraction of a percent above our threshold, and the addition of the TGLC data pushed the noise down so the signal was a fraction of a percent below. The remaining signals were non-transit-like, and would have been dropped during the final human inspection stage. The above shows that the addition of TGLC photometry data to existing data does not significantly affect the planet search.

With each new sector, the ratio of TGLC to QLP-classic data will continue to increase, allowing us to search for smaller planets around fainter stars. We plan to reprocess all TESS data using the new TGLC photometry methods and find signals that are currently too weak to be detected by BLS. We will monitor the effectiveness of \texttt{Astronet-Triage} and retrain as necessary to maximize performance.

In summary, the new TGLC photometry provides QLP light curve users with improved precision across the full range of covered brightness and allows planet searches to go deeper than before. The resulting HLSP FITS files are fully compatible with previous QLP lightcurves.

\section{Acknowledgements}
These data release notes provide processing updates from the Quick-Look Pipeline (QLP) at the TESS Science Office (TSO) at MIT. This work makes use of FFIs calibrated by TESS Image CAlibrator \citep[TICA;][]{TICA}, which are also available as High-Level Science Products (HLSPs) stored on the Mikulski Archive for Space Telescopes (MAST). Funding for the TESS mission is provided by NASA's Science Mission Directorate.

\bibliography{references}
\bibliographystyle{aasjournalv7}


\end{document}